\documentclass[letter]{aa} 
\usepackage{graphicx}
\usepackage{txfonts}
\usepackage{natbib}
\usepackage[breaklinks=false]{hyperref}
\newcommand{\Ms}{{\ensuremath{\mathrm{M}_{\odot}}}}
\newcommand{\Rs}{{\ensuremath{\mathrm{R}_{\odot}}}}
\newcommand{\Ls}{{\ensuremath{\mathrm{L}_{\odot}}}}

\newcommand{\Mpy}{\Ms\,{\rm yr}{\ensuremath{^{-1}}}}
\newcommand{\dM}{\ensuremath{\dot M}}
\newcommand{\gva}{{\sc genec}}
\newcommand{\Ledd}{{\ensuremath{L_{\rm Edd}}}}
\newcommand{\tria}{{\ensuremath{3\alpha}}}

\newcommand{\vits}{{\ensuremath{\rm v_{sound}}}}

\newcommand{\tsc}{{\ensuremath{\tau_{\rm SC}}}}
\newcommand{\tkh}{{\ensuremath{\tau_{\rm KH}}}}
\newcommand{\tac}{{\ensuremath{\tau_{\rm accr}}}}
\newcommand{\Lac}{{\ensuremath{L_{\rm accr}}}}

\usepackage{color}

\begin{document}

\title{Maximally accreting supermassive stars: a fundamental limit imposed by hydrostatic equilibrium}

\author{
L. Haemmerl\'e\inst{\ref{inst1}},
G. Meynet\inst{\ref{inst1}},
L. Mayer\inst{\ref{inst2}},
R. S. Klessen\inst{\ref{inst3},\ref{inst4}},
T. E. Woods\inst{\ref{inst5}},
A. Heger\inst{\ref{inst6},\ref{inst7}}
}
\authorrunning{Haemmerl\'e et al.}

\institute{
D\'epartement d'Astronomie, Universit\'e de Gen\`eve, chemin des Maillettes 51, CH-1290 Versoix, Switzerland        \label{inst1}\and
Center for Theoretical Astrophysics and Cosmology, Institute for Computational Science, University of Zurich, Winterthurerstrasse 190,
CH-8057 Zurich, Switzerland     \label{inst2}\and
Universit\"at Heidelberg, Zentrum f\"ur Astronomie, Institut f\"ur Theoretische Astrophysik, Albert-Ueberle-Str. 2, D-69120 Heidelberg,
Germany \label{inst3}\and
Universit\"{a}t Heidelberg, Interdisziplin\"{a}res Zentrum f\"{u}r Wissenschaftliches Rechnen, Im Neuenheimer Feld 205, D-69120 Heidelberg,
Germany	\label{inst4}\and
National Research Council of Canada, Herzberg Astronomy \& Astrophysics Research Centre, 5071 West Saanich Road, Victoria, BC V9E 2E7, Canada\label{inst5}\and
School of Physics and Astronomy, Monash University, VIC 3800,\,Australia	\label{inst6}\and
Tsung-Dao Lee Institute, Shanghai 200240, China \label{inst7}
}

\date{Received ; accepted }

 
\abstract
{Major mergers of gas-rich galaxies provide promising conditions for the formation of supermassive black holes
(SMBHs; $\gtrsim10^5$ \Ms) by direct collapse because they can trigger mass inflows as high as $10^4-10^5$ \Mpy\ on sub-parsec scales.
However, the channel of SMBH formation in this case, either dark collapse (direct collapse without prior stellar phase)
or supermassive star (SMS; $\gtrsim10^4\,\Ms$), remains unknown.}
{
Here, we investigate the limit in accretion rate up to which stars can maintain hydrostatic equilibrium.
}
{We compute hydrostatic models of SMSs accreting at 1 -- 1000 \Mpy,
and estimate the departures from equilibrium a posteriori by taking into account the finite speed of sound.}
{We find that stars accreting above the atomic cooling limit ($\gtrsim10\,\Mpy$) can only maintain hydrostatic equilibrium once they are supermassive.
In this case, they evolve adiabatically with a hylotropic structure,
that is, entropy is locally conserved and scales with the square root of the mass coordinate.}
{Our results imply that stars can only become supermassive by accretion at the rates of atomically cooled haloes ($\sim0.1-10$ \Mpy).
Once they are supermassive, larger rates are possible.
}
 
 
\maketitle
%

\section{Introduction}
\label{sec-intro}

The formation of the first supermassive black holes (SMBHs) is one of the main problems
in our understanding of the formation of cosmic structures (e.g.~\citealt{rees1978,rees1984,volonteri2010,woods2019}).
The most extreme SMBHs have $M_\bullet\sim10^9$ \Ms\ at redshift $z\sim7$ \citep{mortlock2011,wu2015,banados2018,wang2018}.
Assuming the seed of these objects formed at $z\sim30$, for example through direct collapse,
this value implies the accumulation of mass at average rates of 1 -- 10~\Mpy\ in compact objects for a billion years.
Such rates are found in atomically cooled primordial haloes,
where H$_2$ molecules have been destroyed by an external UV field
(e.g.~\citealt{haiman1997a,omukai2001a,bromm2003b,dijkstra2008,latif2013e,regan2017}).

More recently, an alternative route has been found in the merger of massive galaxies \citep{mayer2010,mayer2015,mayer2019}.
This scenario has two advantages compared to atomically cooled haloes:
first, star formation does not need to be suppressed and second, much larger inflows are obtained, as high as $10^4-10^5$ \Mpy\ at sub-parsec scales.
The gas accumulates on a rotationally supported disc that reaches $\sim10^9$~\Ms;
what happens next however remains unknown.
The disc might either collapse directly to a black hole ({\it dark collapse}, \citealt{mayer2019}),
or  first form a supermassive star (SMS), which then collapses to form a SMBH once accretion stops
or general relativistic (GR) instability sets in (e.g.~\citealt{woods2019}).
Interestingly, the most distant observed quasar seems to be hosted by a galaxy merger \citep{banados2019}
and has an inferred mass similar to that of the massive disc obtained in the simulations.

The properties of SMSs accreting at the rates of atomically cooled haloes (0.1 -- 10 \Mpy)
have been studied both analytically and numerically in the last decade
\citep{begelman2010,hosokawa2012a,schleicher2013,hosokawa2013,sakurai2015,umeda2016,
woods2017,haemmerle2018a,haemmerle2018b,haemmerle2019a}.
They are found to evolve as red supergiant protostars along the Hayashi limit \citep{hosokawa2012a,hosokawa2013},
before collapsing due to the GR instability at several $10^5$~\Ms\ \citep{umeda2016,woods2017,haemmerle2018a}.
The case of galaxy mergers provides conditions for which rates higher than those of atomically cooled haloes are possible.
Here we address the question of whether or not SMSs can form by accretion at rates $\gtrsim100$ \Mpy.
We investigate the conditions in which hydrostatic equilibrium can be sustained for such rates
by computing the corresponding hydrostatic structures numerically and estimating a posteriori the first-order hydrodynamical corrections.
We consider the stability of monolithic SMSs in the context of other formation scenarios in a companion paper (Woods, Heger, \& Haemmerl\'e, in prep.).
For a discussion on the bottlenecks expected in SMS formation, see \citep{woods2019}.

The paper is organised as follows.
In Sect.~\ref{sec-time} we proceed to a preliminary analysis based on global timescales in order to illustrate the main ideas.
In Sect.~\ref{sec-mthd}, we give the numerical method followed to build the stellar models.
The stellar structures are described in Sect.~\ref{sec-rslt} and their implications are discussed in Sect.~\ref{sec-dscs}.
We conclude in Sect.~\ref{sec-out}.

\section{Timescales}
\label{sec-time}

In the absence of nuclear reactions, the evolution of an accreting star is governed by the competition between two processes:
thermal relaxation and accretion.
The first process, appearing mathematically in the energy equation, relies on the variation of entropy $\dot s$.
The second one, appearing in the boundary conditions, relies on the variation of mass \dM.
The efficiency of these two processes can be approximately estimated from their corresponding  global timescales,
the Kelvin-Helmholtz (KH) time and the accretion time:
\begin{equation}
\tkh={GM^2\over RL}
\label{eq-tkh},\end{equation}
\begin{equation}
\tac={M\over\dM},
\label{eq-tac}\end{equation}
where $M$ is the stellar mass, $R$ and $L$ are the photospheric radius and luminosity, respectively,
and $G$ is the gravitational constant.
When $\tac\gg\tkh$, the evolution is essentially governed by thermal contraction
and the star is expected to be thermally relaxed after a KH time, as nuclear reactions are ignited.
When $\tac\ll\tkh$, accretion dominates the evolution, leading to the swelling of the envelope,
because the entropy cannot be radiated away efficiently enough to restore thermal equilibrium
against the perturbative effect of accretion \citep{hosokawa2009}.

In principle, if the accretion time is short enough compared to the KH time (i.e. if accretion is fast enough),
thermal processes become negligible (in terms of structure) and the evolution is adiabatic:
it is not  a loss of entropy that causes the star to contract, but rather the increasing pressure caused by the accreted mass.
This implies that above a given threshold in the accretion rate,
the stellar structure for a given mass no longer depends on the accretion rate.
Indeed, if the evolution is governed by accretion only, the time only appears in the problem through the change in mass.
In the range $\dM\leq10$~\Mpy\ already covered in the literature, thermal processes are still at play
and this pure accretion regime is not reached.

The existence of a pure accretion regime depends on the ability of the star to maintain hydrostatic equilibrium at the required rates.
Mechanical equilibrium in stellar interiors is restored by pressure- and gravity-waves in a sound-crossing time:
\begin{equation}
\tsc={R\over\vits}={R\over\sqrt{\Gamma_1{P\over\rho}}},
\label{eq-tsc}\end{equation}
where \vits\ is the sound-speed, $\Gamma_1$ the first adiabatic exponent,
$P$ the thermal pressure, and $\rho$ the density of mass.
If $\tac\lesssim\tsc$,
the mass increases by a factor of two or more during the time needed to restore hydrostatic equilibrium.
Thus a hydrostatic core cannot incorporate mass at such rates.

\begin{figure}\begin{center}
\includegraphics[width=0.44\textwidth]{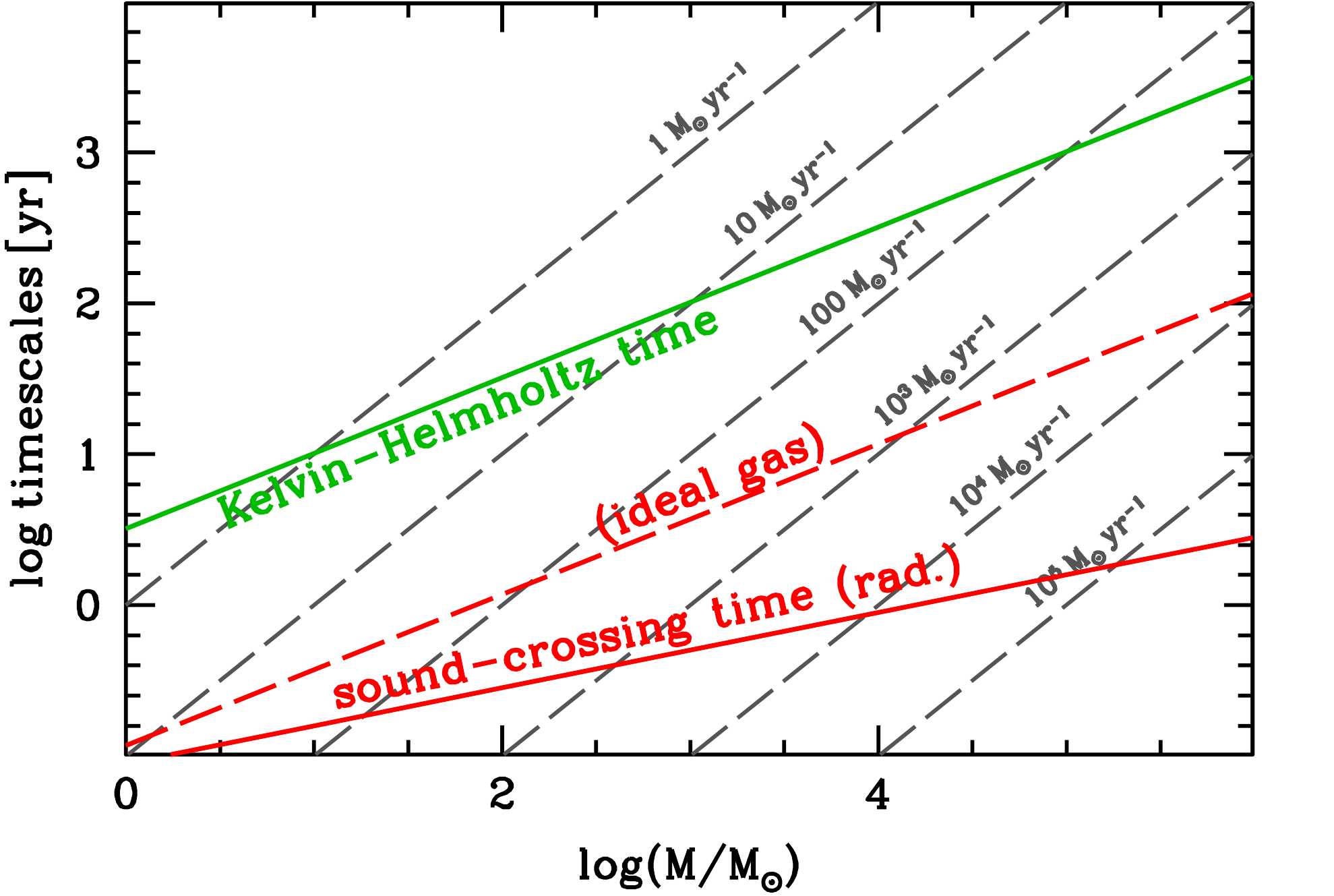}
\caption{Global timescales of Eq.~(\ref{eq-tkh}--\ref{eq-tac}--\ref{eq-tsc}), computed with Eq.~(\ref{eq-fit}).
The sound-crossing time is shown for ideal gas and radiation pressure.
The accretion time is shown for the indicated rates.}
\label{fig-timescales}
\end{center}\end{figure}

Figure~\ref{fig-timescales} shows the timescales of Eq.~(\ref{eq-tkh}--\ref{eq-tac}--\ref{eq-tsc}).
The KH time is obtained using the fits by \cite{hosokawa2012a} for $R$ and $L$:
\begin{equation}
R=260\,\Rs\times\left({M\over\Ms}\right)^{1/2}  \qquad  L=38\,000\,\Ls\times{M\over\Ms}
\label{eq-fit}.\end{equation}
For the sound-crossing time, we estimate the sound speed for cases of ideal gas and radiation pressure.
We assume constant mean molecular weight ($\mu=0.6$) and temperature ($10^5$~K)
relevant for the envelope of accreting SMSs.
For the density, we consider only the outer 10\% of the stellar mass, which covers most of the radius.
For $M>10^4$~\Ms, the sound-crossing time is longer than a year
due to the large radii and low temperatures.
We note the effect of radiation pressure, which dominates in this mass range,
in shortening the sound-crossing time by an order of magnitude.
A comparison with the KH time indicates that a regime of pure accretion might
exist in the high mass range for rates around $10^3-10^4$~\Mpy.

The use of global timescales in stellar evolution is not always conclusive however.
For instance, Fig.~\ref{fig-timescales} suggests that stars of 1000 \Ms\ accreting at 1 \Mpy\ are thermally relaxed,
which is refuted by all numerical models (e.g.~\citealt{woods2019} for a comparison).
For the same reason, a precise estimate of the maximal accretion rate that allows for hydrostatic equilibrium requires the consideration of numerical models.

\section{Numerical method}
\label{sec-mthd}

The numerical models are computed with the same method as \cite{haemmerle2018a}, that is, with \gva,
a one-dimensional hydrostatic stellar evolution code that solves the equations of stellar structure numerically.
Detailed descriptions of the code are available in the literature (e.g.~\citealt{eggenberger2008}).
In the present case, the central ingredient is accretion \citep{haemmerle2016a}, included through cold accretion;
that is the entropy of the accreted gas matches that of the stellar surface \citep{palla1992}.
For rates of atomically cooled haloes, the effect of hot accretion (i.e. advection of entropy; \citealt{hosokawa2009})
can be neglected since the intrinsic luminosity, which is nearly Eddington,
dominates the energy budget at the stellar surface \citep{hosokawa2013}.
For larger rates, this assumption might not be justified, which is discussed in Sect.~\ref{sec-hylo}.
Rotation is not included,
since accreting SMSs are found to be slow rotators due to the $\Omega\Gamma$-limit,
which implies negligible impact of rotation on their structure \citep{haemmerle2018b}.

For zero metallicity, models with $\dM\leq10$ \Mpy\ are already described in \cite{haemmerle2018a}.
Models of rapidly accreting SMSs at solar metallicity have not yet been published. 
For the present study, we compute four new models.
We first extend the set of Pop III models by considering the rates 100~\Mpy\ and 1000 \Mpy.
In order to understand the impact of metals in various regimes,
we compute two models at solar metallicity with rates 1 and 1000 \Mpy.
We highlight the fact  that we do not include any mass-loss, even at solar metallicity, due to the short evolutionary timescales.
We use initial models of 5 and 10 \Ms\ for numerical stability, with fully convective structures similar to those used in \cite{haemmerle2018a}.
The models are run until the limit of numerical stability, which is suspected to be related to pulsation instability.

\section{Results: hydrostatic structures}
\label{sec-rslt}

The internal structures and surface properties of the new models are qualitatively similar to those of the models already published.
They evolve along the Hayashi limit and start to burn H in their core at the latest when they reach $10^5$ \Ms.
Hydrogen burning triggers convection in the core, but most of the mass remains radiative.
Hydrostatic objects forming at rates $\lesssim1000$ \Mpy\ go through energetically significant nuclear reactions,
and are thus stars, sensu stricto.

The main properties of the models are summarised in Tables \ref{tab-fin}, \ref{tab-burn}, and \ref{tab-core}.
The highest mass we reach is 800 000 \Ms\ for the model of 1000 \Mpy\ at solar metallicity.
The corresponding lifetime is as short as 800 yr.
The mass at which H starts to burn increases with the accretion rate until this latter reaches at least 100 -- 1000 \Mpy\ .
However, the mass fraction of the convective core at a given mass depends weakly on the rate above 100 \Mpy.
For $Z=0$, this fraction remains always below 1\%; at solar metallicity, it grows to at most 8\% at the end of the run.

\begin{table}
\caption{Masses at the end of the run (set by numerical instability) in $10^5$~\Ms\  for the models with various accretion rates and metallicities.
The new models are marked in boldface.}
\label{tab-fin}
\centering
\begin{tabular}{c|cccc|}
                        & 1 \Mpy                                & 10 \Mpy                               & 100 \Mpy                & 1000 \Mpy             \\
\hline
$Z=0$           & 2.29  & 5.43  &{\bf4.25}      &{\bf5.25}      \\
$Z=Z_\odot$     &{\bf3.02}      &               &               &{\bf8.57}      \\
\hline
\end{tabular}
\end{table}

\begin{table}
\caption{Stellar mass at the beginning of H-burning.}
\label{tab-burn}
\centering
\begin{tabular}{c|cccc|}
                        & 1 \Mpy                & 10 \Mpy               & 100 \Mpy                & 1000 \Mpy             \\
\hline
$Z=0$           & 5000 \Ms      & 20 000 \Ms    &{\bf50 000 \Ms}        &{\bf100 000 \Ms}        \\
$Z=Z_\odot$     &{\bf2000 \Ms}  &                       &                               &{\bf40 000 \Ms}        \\
\hline
\end{tabular}
\end{table}

\begin{table}
\caption{Mass fraction of the convective core at a stellar mass of 200 000 \Ms.}
\label{tab-core}
\centering
\begin{tabular}{c|cccc|}
                        & 1 \Mpy                & 10 \Mpy       & 100 \Mpy      & 1000 \Mpy       \\
\hline
$Z=0$           & 20\%          & 5\%   &{\bf0.5\%}     &{\bf0.4\%}     \\
$Z=Z_\odot$     &{\bf35\%}      &               &                       &{\bf1\%}               \\
\hline
\end{tabular}
\end{table}

\begin{figure}\begin{center}
\includegraphics[width=0.39\textwidth]{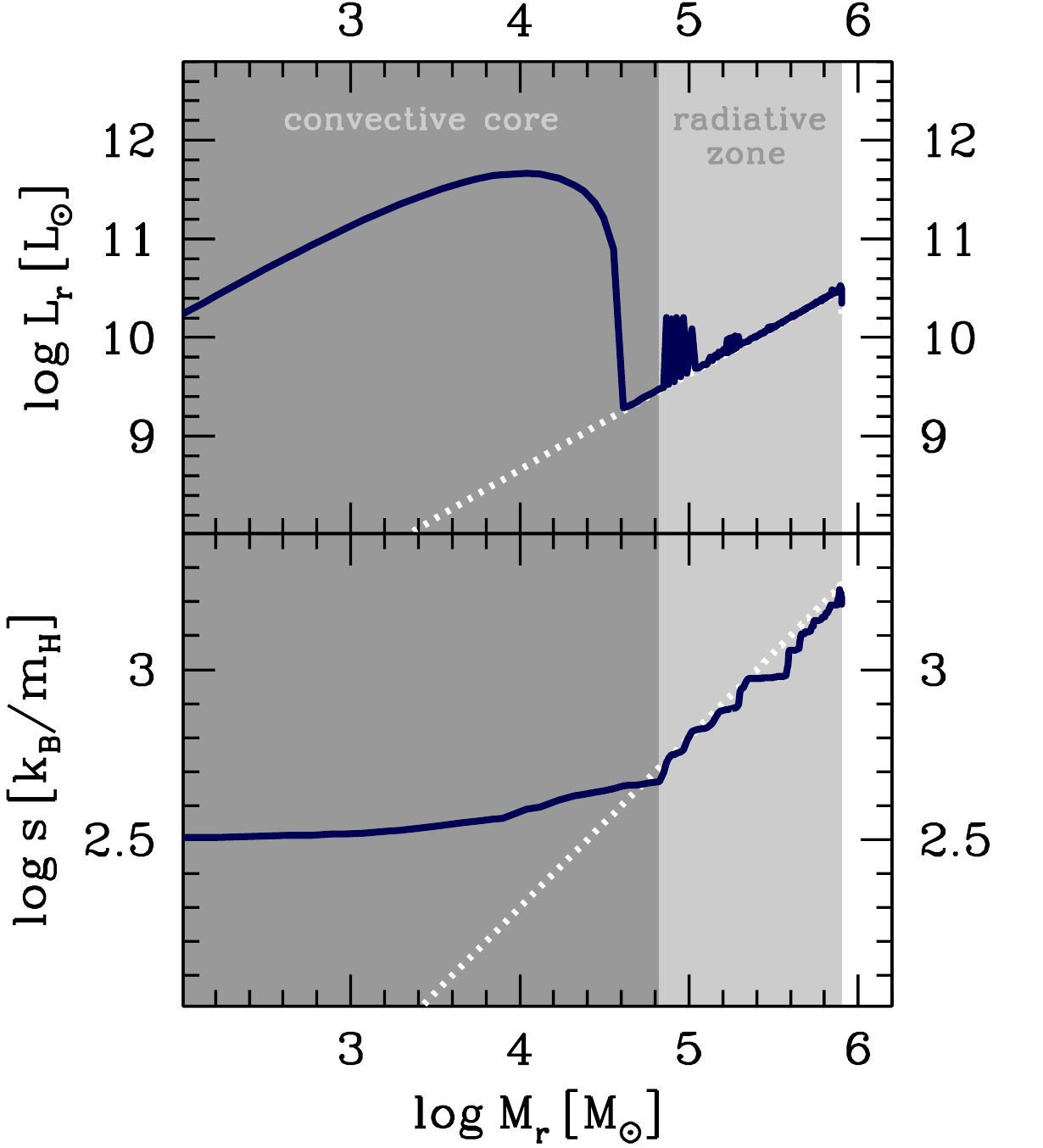}
\caption{Internal profiles of the model at $\dM=1000$ \Mpy, $Z=Z_\odot$ and $M=8\times10^5$ \Ms.
Upper panel: Luminosity profile (solid blue) and Eddington luminosity (dotted white).
Lower panel: Entropy profile (solid blue) and hylotropic fit $s\propto M_r^{1/2}$ (dotted white).}
\label{fig-profils}
\end{center}\end{figure}

The typical structure of a star accreting at 1000 \Mpy\ is illustrated in Fig.~\ref{fig-profils},
with the internal profiles of luminosity and entropy of the model at solar metallicity,
taken near the endpoint of the run, at a mass of $8\times10^5$ \Ms.
The convective core contains 8\% of the total stellar mass.
Most of the mass is in the radiative zone, where the luminosity profile matches the local Eddington luminosity,
$\Ledd(r)=4\pi cGM_r/\kappa$ ($c$ is the speed of light and $\kappa$ the opacity given by the model, which is
dominated by electron-scattering),
and the entropy profile follows a power law $s\propto M_r^{1/2}$.
This power law corresponds to the hylotropic structures of \cite{begelman2010}.
The entropy profiles at successive stages match each other in the radiative region, indicating negligible entropy losses,
that is, a pure accretion regime in most of the star.

Of course, this is not the case in the convective core,
where the entropy released by H-burning is so huge that the local luminosity
exceeds the local Eddington luminosity by three orders of magnitude,
and even the surface Eddington limit by one order of magnitude (Fig.~\ref{fig-profils}, upper panel).
We note that the radiative luminosity is never super-Eddington in the core,
meaning that density or temperature inversions are not required.
Thermal processes are so inefficient that the released nuclear entropy cannot be transported by radiation
and remains captured in the convective layers.
As a consequence, the luminosity drops by several orders of magnitude at the boundary of the core,
and converges rapidly to the Eddington profile.
It follows that the evolution of the envelope is decoupled from that of the core.
In particular, the energy radiated at the surface arises from the thermal contraction of the envelope,
not from central H-burning.
This is in spite of the fact that thermal processes have negligible impact on the structure.
Indeed, the dominating process (adiabatic accretion-driven contraction) does not release any entropy,
and thus does not contribute to the luminosity.

The impact of metals on stellar structures is twofold: they increase the opacity and allow for the CNO cycle to operate without previous \tria\ reactions.
In the present case, the effect of high opacity is to lock the star on the Hayashi limit for longer,
avoiding the drift towards the blue, as obtained for $\gtrsim10^9$ \Ls\ in previous studies at zero metallicity \citep{hosokawa2013,haemmerle2018a}.
This is due to the large number of free electrons provided by the metals that are available for H$^-$ formation,
whose opacity law keeps the photospheric temperature constant.
This fact implies a weaker UV feedback in the supermassive range compared to zero metallicity.
In solar metallicity models, the possibility for the CNO cycle to start before \tria\ reactions
implies that H-burning is triggered at lower central temperatures, that is, at lower masses (Table~\ref{tab-burn}).
The main consequence of this is an increase in the mass fraction of the convective core for a given total mass
(Table~\ref{tab-core}) since entropy starts to be released earlier.
Another consequence is a lower central temperature and density for a given mass,
since the pre-H-burning contraction is stopped earlier.

The final collapse of SMSs relies on the GR instability, which is a radial pulsation instability
and cannot be captured by hydrostatic models.
Polytropic models provide a simple criterion \citep{chandrasekhar1964} however,
which is not conclusive in the present case.
This criterion might apply only in the core, which corresponds to a polytrope.
For rates of atomically cooled haloes, it fails to predict correct final masses,
but still allows the trend of the dependence on the accretion rate to be captured
\citep{woods2017,haemmerle2018a,woods2019}.
The final masses given by the polytropic criterion in the present case
remain in the range of those of previous studies at lower rates,
which confirms that above 0.1~\Mpy\ the final mass depends weakly on the rate.
For $Z=0$, the instability arises at $\sim5\times10^5$ \Ms\ in the models at 100 -- 1000 \Mpy.
The models at solar metallicity have larger cores but also lower densities (i.e. lower GR correction),
which can delay the instability by a factor of approximately two in mass and age.
Overall, if we consider the polytropic criterion in the core only,
the most massive stable model we obtain is for 1000~\Mpy\ and $Z=Z_\odot$, and has a mass of $6.8\times10^5$ \Ms.

\section{Discussion}
\label{sec-dscs}

\subsection{Hydrostatic equilibrium}
\label{sec-hydro}

As discussed in Sect.~\ref{sec-time}, hydrostatic equilibrium requires that evolution proceed slowly enough
for the pressure- and gravity waves to cross the star before it evolves significantly.
Since for the rates considered here the accretion time becomes similar to the sound-crossing time,
the consistency of the models must be questioned.

Hydrodynamical effects become important when the acceleration term $\ddot r$ in the momentum equation,
\begin{equation}
\ddot r=-{1\over\rho}\nabla P-g\,,
\label{eq-dyn}\end{equation}
becomes comparable with the other terms, the gravitational acceleration $g>0$
and the acceleration due to the pressure gradient $\nabla P$.
In order to estimate the departures from hydrostatic equilibrium,
we derive hydrodynamical structures from the hydrostatic structures described in Sect.~\ref{sec-rslt},
by assuming each layer $M_r$ has, at each age $t$, the properties ($P,\rho,g$)
of the hydrostatic structure at a previous age $t'$,
with a time-delay $t-t'$ corresponding to the time it takes for a sound wave to cross the layers above $M_r$.
Modifications in the boundary conditions (i.e. accretion) during this delay cannot impact the layers below $M_r$
because this would require supersonic transport of momentum.
Since the evolution is driven by accretion, a layer has no reason to evolve except under the effect of accretion.
The time-delay is computed by integration of the inverse of the sound speed (denominator of Eq.~\ref{eq-tsc})
over a path comoving with the contracting gas.
Assuming the modified profiles of $P$, $\rho,$ and $g$ mimic the hydrodynamical structures at first order,
we can derive the acceleration term with Eq.~(\ref{eq-dyn}).
We note that $\nabla P$ is re-derived from the new profiles: adjacent layers are delayed to different hydrostatic models,
which induces a correction to $\nabla P$ at the origin of the non-zero acceleration term in Eq.~(\ref{eq-dyn}).
The condition for hydrostatic equilibrium is then obtained by comparing this acceleration term with the gravitational term.
We assume hydrodynamical effects become significant when $\vert\ddot r\vert\gtrsim0.1g$.
This method does not allow the actual hydrodynamical evolution to be captured however,
but the underlying idea is the following:
if, even when accounting for the time it takes for the various layers to come under the effects of accretion,
the corrected structures do not induce significant acceleration, and the evolution remains hydrostatic.

The larger the stellar mass, the weaker the impact of accretion of a given amount of mass on the structure.
Therefore, for a constant accretion rate the evolution slows down as the mass grows,
which is reflected by the increase of the accretion time with mass (Eq.~\ref{eq-tac}, Fig.~\ref{fig-timescales}).
As a consequence, the star reaches stability more easily as its mass grows,
and the condition for hydrostatic equilibrium corresponds to a minimum mass for any given rate.
Moreover, due to their low temperatures, large radii, and rapid contraction, the outer layers are the least stable,
meaning that hydrostatic equilibrium is broken at higher masses in the envelope than in the core.
Thus for each accretion rate, we derive the minimum mass at which the whole star becomes hydrostatic
as well as that at which the core only can stabilise.

\begin{figure}\begin{center}
\includegraphics[width=0.44\textwidth]{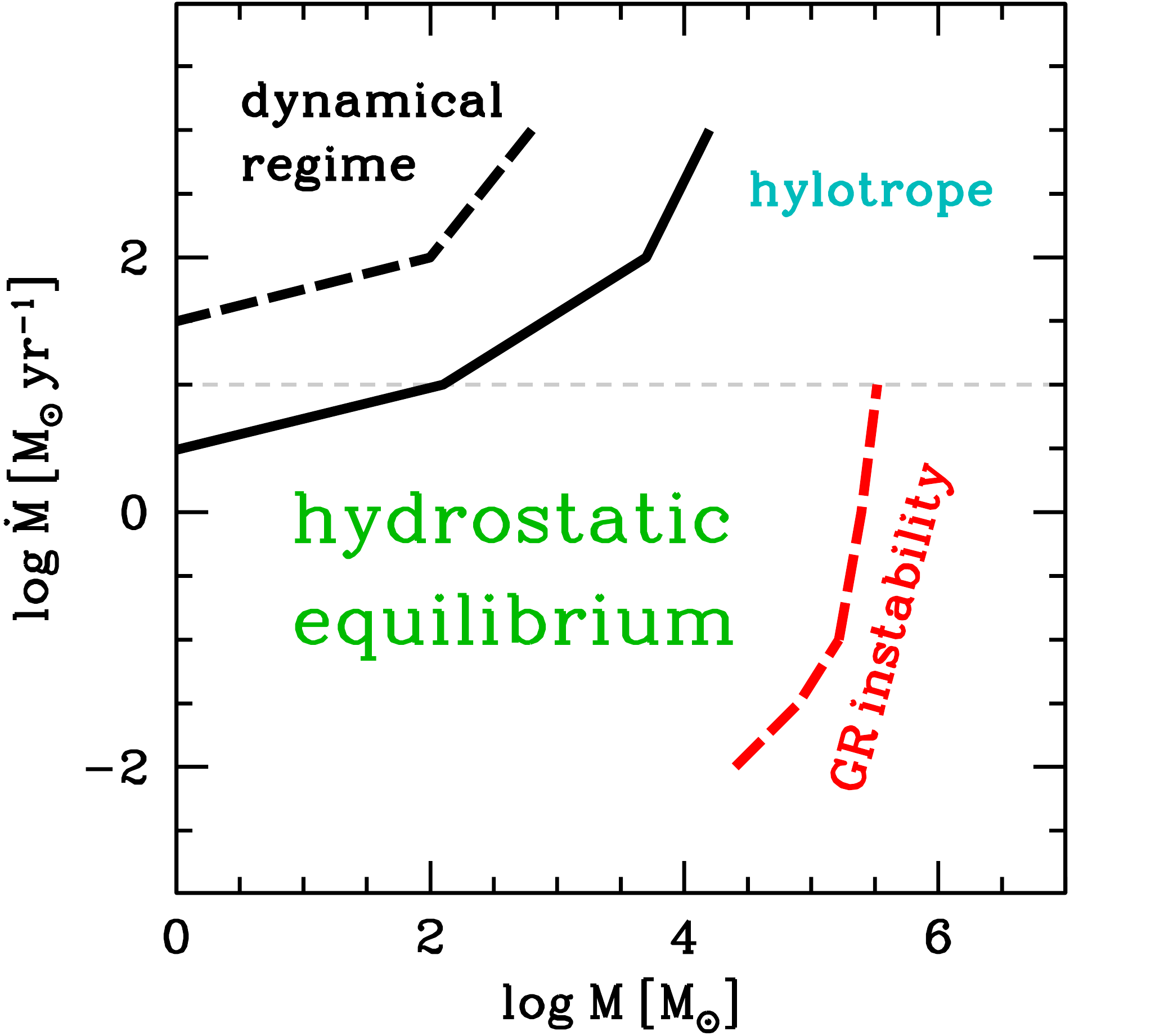}
\caption{Limits to hydrostatic equilibrium for accretion above $10^{-3}$ \Mpy\ (without rotation).
Sound waves can maintain full hydrostatic equilibrium only below the solid black line.
Above the dashed black line, the whole mass evolves hydrodynamically.
Stars become GR unstable at the right-hand side of the red dashed line.
The maximum rate of atomically cooled haloes is shown by a grey dashed line.
The location of the hylotropic, pure-accretion regime is indicated.}
\label{fig-mmdot}
\end{center}
\end{figure}

Figure~\ref{fig-mmdot} summarises the limits to hydrostatic equilibrium for the case at zero metallicity.
The models at solar metallicity do not exhibit significant differences.
The limit to equilibrium arising from accretion is shown by the two black curves,
between which only a fraction of the star can stabilise.
Departures from equilibrium remain negligible for rates $\leq1$ \Mpy,
and the model at 10 \Mpy\ becomes stable as soon as it exceeds 100~\Ms.
Hydrodynamical effects become significant for rates $\geq100$ \Mpy.
The core can only stabilise above a stellar mass of 100 -- 1000 \Ms,
and the star as a whole above $\gtrsim10^4$ \Ms.
Therefore, accretion at rates $\gtrsim100$ \Mpy\ is
only compatible with mechanical equilibrium for stars more massive than $10^4$ \Ms.
The red line on Fig.~\ref{fig-mmdot} shows the limits to equilibrium resulting from GR instability
obtained by \cite{woods2017}.
An extrapolation of the curve (see also Sect.~\ref{sec-rslt}) suggests that a hylotropic,
pure-accretion regime can be stable over one or two orders of magnitude in mass.
Additional effects, such as rotation, could extend the mass range further by stabilising the star against the GR instability
\citep{fowler1966,bisnovatyi1967}.

\subsection{Hylotropic structures}
\label{sec-hylo}

Hylotropic models have been developed by \cite{begelman2010} to analytically study the structures of accreting SMSs.
These were applied to the quasi-star model \citep{begelman2008},
where a hydrostatic envelope is supported by the energy of accretion onto a central BH.
The stability of such systems requires the BH to be $\lesssim1\%$ of the envelope mass.
The question remains open as to whether or not accreting SMSs collapsing through the GR instability can form such objects
\citep{hosokawa2013,woods2019}.
Since the GR instability is thought to appear in the core first, a core of $\lesssim1\%$ of the total mass is probably required.
The models of SMSs accreting at the rates of atomically cooled haloes have larger cores ($\gtrsim10\%$~M),
but interestingly in the models of the present study at 100 -- 1000 \Mpy\
and $Z=0$ the core mass has the required fraction, while the rest of the star corresponds indeed to a hylotrope.

We note however that for the rates required for hylotropic structures,
the gravitational energy liberated by accretion becomes similar or exceeds the energy radiated at the photosphere.
Indeed, with the fits of Eq.~(\ref{eq-fit}), the ratio of the accretion luminosity $\Lac=GM\dM/R$ to the intrinsic luminosity is
\begin{equation}
{\Lac\over L}\sim\left({M\over10^6\,\Ms}\right)^{-1/2}{\dM\over100\,\Mpy},
\end{equation}
which shows that hot accretion (Sect.~\ref{sec-mthd}) becomes dominant
if the rate exceeds 100 \Mpy\ before the mass reaches $10^6$~\Ms.
Nevertheless, assuming that the heat advected per mass unit, $T\Delta s_{\rm hot}$,
scales with the specific energy of free-fall, $GM/R$, and that the temperature at accretion is constant (Hayashi limit),
the fit of Eq.~(\ref{eq-fit}) for $R$ implies that $\Delta s_{\rm hot}\propto M_r^{1/2}$, that is, the structure remains hylotropic.

\section{Summary and conclusions}
\label{sec-out}

The models of the present study extend the previous works on rapidly accreting SMSs
by considering larger rates (100 -- 1000~\Mpy)
and solar chemical composition.
The new stellar structures display most of the main features of the previous models:
the star evolves as a red supergiant protostar with most of the mass being radiative,
and burns hydrogen in a convective core before GR instability leads to collapse, forming  a SMBH.
The polytropic criterion indicates an onset of GR instability at masses of $10^5-10^6$ \Ms.

The main quantitative differences are in the size of the convective core,
which represents only $\lesssim1\%$ of the total stellar mass for the models at 100 -- 1000 \Mpy\ and zero metallicity.
Adding metals at solar abundances changes the core mass fraction by a factor of a few.
For rates above the atomic cooling limit ($\gtrsim10-100$~\Mpy),
thermal processes become inefficient and the evolution is governed by accretion only, leading to hylotropic structures.

However, the evolution at $\gtrsim100$ \Mpy\ is so fast that hydrostatic equilibrium
can only be sustained if the star is already supermassive.
At lower masses, pressure- and gravity-waves cannot restore hydrostatic equilibrium fast enough.
This implies that stars can become supermassive by accretion only at the rates of atomically cooled haloes.
Once they are supermassive, larger rates are possible.
This result implies that forming hylotropic stars by accretion requires rates that increase with time.
The evolution in the dynamical accretion regime, in particular the possibility for dark collapse,
shall be addressed in a future study.

\begin{acknowledgements}
This work was sponsored by the Swiss National Science Foundation (project number 200020-172505).
RSK acknowledges financial support from the German Science Foundation (DFG) via the collaborative research centre (SFB 881) “The Milky Way System” (subprojects B1, B2, and B8) and from the Heidelberg cluster of excellence EXC 2181 "STRUCTURES: A unifying approach to emergent phenomena in the physical world, mathematics, and complex data'' funded by the German Excellence Strategy.
TEW acknowledges support from the NRC-Canada Plaskett Fellowship.
We thank Kazuyuki Omukai and Takashi Hosokawa for fruitful discussions.
\end{acknowledgements}

\bibliographystyle{aa}
\bibliography{bib}

\end{document}